\documentclass[%
 aip,
 amsmath,amssymb,
preprint,%
author-year,%
]{revtex4-1}

\usepackage{graphicx}% Include figure files
\usepackage{dcolumn}% Align table columns on decimal point
\usepackage{bm}% bold math
\usepackage{xcolor}
\usepackage[utf8]{inputenc}
\usepackage[T1]{fontenc}
\usepackage{mathptmx}
\usepackage{etoolbox}
\usepackage{caption}
\usepackage{subcaption}
\usepackage{tikz}

\makeatletter
\def\@email#1#2{%
 \endgroup
 \patchcmd{\titleblock@produce}
  {\frontmatter@RRAPformat}
  {\frontmatter@RRAPformat{\produce@RRAP{*#1\href{mailto:#2}{#2}}}\frontmatter@RRAPformat}
  {}{}
}%
\makeatother
\begin{document}

\preprint{AIP/123-QED}

\title[]{A constitutive model for viscosity of dense fiber suspension.}%:\\with Forced Linebreak}
% Force line breaks with \\
\author{A. Monsurul Khan}
 \affiliation{Department of Mechanical Engineering, Purdue University, IN 47905, USA }%Lines break automatically or can be forced with \\
\author{B. Rishabh V. More}%
 
\affiliation{ 
Department of Mechanical Engineering, Massachusetts Institute of Technology, MA 02139, USA%\\This line break forced with \textbackslash\textbackslash
}%

\author{Arezoo M. Ardekani}
\email{ardekani@purdue.edu.}
% \homepage{http://www.Second.institution.edu/~Charlie.Author.}
\affiliation{%
Department of Mechanical Engineering, Purdue University, IN 47905, USA%\\This line break forced% with \\
}%

\date{\today}% It is always \today, today,

\begin{abstract}\

We propose a constitutive model to predict the viscosity of fiber suspensions, which undergoes shear thinning, at various volume fractions, aspect ratios, and shear stresses/rates. We calibrate the model using the data from direct numerical simulation and prove the accuracy by predicting experimental measurements from the literature. We use a friction coefficient decreasing with the normal load between the fibers to quantitatively reproduce the experimentally observed shear thinning in fiber suspensions.
%We establish the accuracy and effectiveness of the model by using a large data set obtained from direct numerical simulations of the shear flow of fiber suspensions with a load-dependent coefficient of friction between the fibers to physically capture the shear-thinning behavior of these suspensions. %\textcolor{blue}{These suspensions show increasingly strong shear thinning as the fiber aspect ratio as well as the solid volume fraction increases.}
%In dense suspensions, the fibers come into contact due to surface asperities. As a result, contact force and inter-fiber friction become the dominant factors governing the rheology.
%A model for this experimentally observed shear-thinning viscosity is provided by the comparison of simulations performed with either a constant or a load-dependent friction coefficient. 
In this model, the effective normal contact force, which is directly proportional to the bulk shear stress, determines the effective friction coefficient. A rise in the shear stress reduces the effective friction coefficient in the suspension. As a result, the jamming volume fraction increases with the shear stress, resulting in a shear thinning in the suspension viscosity. Moreover, we extend the model to quantify the effects of  fiber volume fraction and aspect ratio in the suspension.  We calibrate this model using the data from numerical simulations for the rate-controlled shear flow. Once calibrated, we show that the model can be used to predict the relative viscosity for different volume fractions, shear stresses, and aspect ratios. The model predictions are in excellent agreement with the available experimental measurements from the literature. The findings of this study can potentially be used to tune the fiber size and volume fraction for designing the suspension rheology in various applications.
%In this study, the behavior is simulated over wide ranges of volume fraction, aspect ratio,  dimensionless shear stress, and coefficient of inter-fiber friction. 
\end{abstract}

\maketitle

%\begin{quotation}
%The ``lead paragraph'' is encapsulated with the \LaTeX\ 
%\verb+quotation+ environment and is formatted as a single paragraph before the first section heading. 
%(The \verb+quotation+ environment reverts to its usual meaning after the first sectioning command.) 
%Note that numbered references are allowed in the lead paragraph.
%
%The lead paragraph will only be found in an article being prepared for the journal \textit{Chaos}.
%\end{quotation}
\section{\label{sec:introduction}Introduction:\protect\\ }
Suspensions of fibers are found in a wide variety of applications, including mixing of elongated particles with concrete to enhance strength, tailoring the rheological properties of drilling fluids,  and the production of paper from wood \citep{bivins2005new, lundell2011fluid, hassanpour2012lightweight, elgaddafi2012settling}. During processing and transport, these materials are subjected to shear deformations, which cause translation, rotation, bending, and breaking of the fibers. These phenomena modify the microstructure in the suspension, which in turn affects the final product's physical and mechanical properties. Therefore, it is desirable to have an accurate prediction of the rheological characteristics of suspensions in order to optimize these processes in industrial facilities. However, rigid fiber suspensions exhibit a complex rheological behavior due to the gamut of determining variables involved, including fiber-fiber, fiber-wall, and fiber-matrix interactions, as well as events such as fiber breakage and migration. The complex rheological behavior includes non-Newtonian flow properties such as shear thinning, finite normal stress differences, yielding, and jamming \citep{goto1986flow, kitano1981rheology, bounoua2016shear,snook2014normal,keshtkar2009rheological,tapia2017rheology}, making their flow challenging to predict and control. %Shear thinning in fiber suspension is a phenomenon in which the viscosity decreases, sometimes by order of magnitude or more for the applied shear rate \citep{bounoua2016shear}. There are various phenomenological explanations behind shear thinning in fiber suspensions, such as the increase in the effective particle size due to the presence of electric double layer \citep{quemada2002energy}, excluded volume interactions between rigid fibers \citep{raghavan2012conundrum}, elastic bending of flexible micron-sized or nano-sized fibers \citep{bennington1990yield,song2005influence}, fiber aggregation \citep{ma2008rheological} and nonlinear lubrication force \citep{natale2014rheological}. However, it is not possible to create a model or make quantitative predictions based on these phenomenological explanations. Recently adhesive force was attributed to the shear thinning rheology for colloidal suspension but this explanation will not hold for finite size fibers \citep{bounoua2016apparent}.

One of the crucial variables governing suspension rheology is fiber orientation. Hence, most of the constitutive modeling efforts to date have focused on explaining the suspension viscosity by linking fiber orientations to the suspension viscosity. Experimental measurements and theoretical analyses have shown that a higher fiber alignment with the flow results in a lower viscosity when compared to random orientation states. 
A close approximation of the motion of the fiber is frequently achieved by applying Jeffery's exact solution to the motion of an isolated ellipsoid in an infinite Newtonian fluid \citep{jeffery1922motion}. \cite{folgar1984orientation} developed the widely used variation of Jeffery's equations based on rotational diffusion and allowed for the influence of interactions on orientation. These tests from \cite{folgar1984orientation}, Stover \citep{stover1992observations}, and Petrich \citep{petrich2000experimental} conducted on the distribution of fiber orientation demonstrated that an increase in the fiber concentration caused a modest change in the orientation of the fibers during steady shear towards the flow-velocity-gradient plane, which led to an increase in the suspension's viscosity as a whole. Originated by  \cite{hinch1975constitutive,hinch1976constitutive}, the constitutive equation of fiber orientation-induced extra stress in a Newtonian viscous matrix was developed by Dinh and Armstrong \citep{dinh1981rheology,dinh1984rheological}. Furthermore, Dinh-Armstrong's constitutive model has been used to predict the transient shear viscosity \citep{sepehr2004rheological,eberle2009using}; the change of fiber orientation was determined with a strain reduction factor of slowing down the orientation response \citep{folgar1984orientation, advani1990numerical,wang2008objective}. Moreover, Dinh-Armstrong's model has been extended in nonlinear Newtonian viscous liquids, including the Carreau model \citep{ferec2009modeling}, the power law model \citep{ferec2016effect}, and the Bingham model \citep{ferec2017rheological}. The majority of the previous studies have focused on the dilute and semi-dilute regime; however, concentrated fiber suspensions have not been studied as much \citep{batchelor1971stress,goddard1976stress,goddard1976tensile,shaqfeh1990hydrodynamic}. In concentrated suspension, direct interactions between fibers are typical. So, the constitutive model constructed for dilute/semi-dilute suspension must be modified, or new models must be reconstructed \citep{babkin1989constitutive} for the concentrated suspensions. Pipes and coworkers constructed the non-Newtonian constitutive relationships for hyper-concentrated fiber suspensions with an oriented fiber assembly. However, their model is only limited to concentrated suspensions with fibers of very large aspect ratio (>100), where the fibers are required to be arranged with a very high degree of collimation \citep{pipes1991constitutive,pipes1992anisotropic,pipes1994non}. %Moreover, these theories and simulations do not account for hydrodynamic interactions. The simulations with pure mechanical contacts do not predict shear-thinning rheology \citep{sundararajakumar1997structure}, hinting at the lack of a complete physical understanding of micro-mechanics of fiber suspensions. 
Moreover, the studies that only consider hydrodynamic interactions or pure mechanical contacts \citep{sundararajakumar1997structure} cannot predict shear-thinning rheology, hinting at the need for a complete physical understanding of micro-mechanics of fiber suspensions.

Inter-fiber interactions between nearby fibers play a significant role in determining the stress in dense fiber suspensions \citep{sundararajakumar1997structure,lindstrom2007simulation,khan2021rheology}. This is due to the fact that contacts produce friction and resistance to rolling motion in addition to generating normal forces between the fibers. As the contribution of 
frictional contacts to the suspension stress becomes dominant in concentrated suspensions, it is crucial to utilize an accurate contact model that can capture their rate-dependent rheological behavior. However, the simplified model of a constant friction coefficient employed in earlier studies may not be applicable to real fibers and hence, cannot accurately predict the rate-dependence of suspension rheology \citep{salahuddin2013study,lindstrom2008simulation}. Experimental measurements
reveal that the coefficient of friction is not a constant and depends on the normal force \citep{brizmer2007elastic,lobry2019shear,khan2021rheology}. Using the load-dependent friction coefficient model, it has been shown that we can quantitatively predict the shear thinning rheology \citep{khan2021rheology}. 

%In this paper, we focus on developing a constitutive model for the shear thinning rheology. The proposed model is based on two diverging stress-independent rheology, where the properties in between two extremes can be interpolated by a function of stress.

%The nature of interactions between fibers primarily depends on volume fraction and aspect ratio of fibers that determines the suspension viscosity. 

In many industrial applications, carrying suspensions at high solid volume fractions is essential to maximize transportation and reduce energy consumption. Prior rheological research mainly dealt with suspensions at relatively small volume fractions \citep{folgar1984orientation, advani1990numerical,petrie1999rheology,bibbo1987rheology,chaouche2001rheology,salahuddin2013study}. The prediction of Phan \textit{et al.} \citep{phan1991flow}  agrees reasonably well with the Miliken \textit{et al.} \citep{forth1989milliken} data of relative viscosity at a volume fraction less than 10\%. At moderately higher volume fractions, it was observed that the specific viscosity increased with the cube of the volume fraction. The constitutive model for dilute suspensions \citep{giesekus1962elasto, leal1971effect, hinch1973time} can only predict that the specific viscosity is proportional to the volume fraction; the transition from linear to cubic behavior in the relative viscosity vs. the volume fraction is beyond the scope of the dilute-suspension theory. Consequently, this constitutive model for dilute suspensions was extended to the semi-dilute regime to capture the suspension's transition from linear to cubic behavior \citep{phan1991new}. However, to the best of the author's knowledge, there has been no constitutive  model to capture the relative viscosity of the  concentrated suspensions. Furthermore, for fibers with large aspect ratios ($AR=L/d$, where $L$ and $d$ are the fiber length and diameter, respectively), it
is challenging to identify rheology measurements for volume fractions above 0.10. Measurements are only available for volume fractions up to $\phi = 0.15$ or $\phi = 0.17$ \citep{bibbo1987rheology,bounoua2016apparent}, even for aspect ratios as high as 17 or 18, while Bibbó \textit{et al.} \citep{bibbo1987rheology} made measurements as high as $\phi = 0.23$ for smaller aspect ratios of $AR = 9$. Additionally, the volume fraction for non-colloidal fibers at which the shear stresses diverge, and the flow of the suspension ceases (i.e.,
jams) was only determined for a lower aspect ratio, $AR = 14$ \citep{tapia2017rheology}.
Therefore, it is challenging to describe and predict the rheological characteristics of suspensions close to jamming at different aspect ratios due to the lack of a model accurately capturing the underlying physics.

This paper aims to resolve these limitations by quantifying the effect of increasing volume fraction, coefficient of friction, and  fiber aspect ratio on the rheology of relatively rigid fiber suspensions using constitutive equations that can accurately capture these effects. We perform extensive numerical simulations by varying
the volume fraction, aspect ratio, coefficient of friction, and shear rate. Informed by the numerical results, we propose a viscosity model that expresses the suspension rheology in terms of the shear stress and the fiber aspect
ratio. The proposed model is based on two diverging stress-
independent rheological behaviors, where a function of stress can interpolate the properties between two extremes. To this end, we briefly discuss the governing equations, inter-fiber interactions, and simulation conditions in Sec. \ref{sec:methodology}. We then illustrate the constitutive model in Sec.  \ref{sec:model}. Finally, in Sec. \ref{sec:result}, we demonstrate the efficacy of the model by applying it to our simulation data in predicting the suspension rheology. The model
requires the knowledge of rheological data at low and high shear limits
along with an interpolating function of applied shear stress to
predict the relative viscosity at intermediate stress values.

\section{\label{sec:methodology}Methodology:\protect\\ }
This section describes the models and algorithms used to simulate the shear flow of the suspension of fibers. We consider a suspension of N fibers of aspect ratio $AR$ in a shear flow with top and bottom walls moving in the opposite direction with imposing a shear rate $\dot{\gamma}$. We have used the same numerical method to simulate the dense fiber suspensions as in our previous study \citep{khan2021rheology}. %A schematic diagram of the computational configuration and co-ordinate system is shown in  figure~\ref{fig:geometry}.

\subsection{Governing equations}
As the fibers are suspended in a fluid flow, the hydrodynamic forces acting on the fibers need to be related to their deformation to get their configuration in the flow. Thus, the elasticity equation of slender bodies is solved. Next, the governing equations for fluid flow and motion of flexible fibers are introduced.
\subsubsection{Fluid flow}

The suspending fluid is considered an incompressible viscous fluid  with a constant density that is governed by the Navier-stokes equations and the continuity equation.  
\begin{equation}
\frac{{\partial \mathbf{u}}}{{\partial t}} + \mathbf{\nabla} \cdot (\mathbf{u} \otimes \mathbf{u}) =  - \nabla p + \frac{1}{{{\mathop{\rm \textit{Re}}\nolimits} }}{\mathbf{\nabla} ^2}\mathbf{u} + \mathbf{f},
\label{NS}
\end{equation}
\begin{equation}
\mathbf{\nabla} \cdot \mathbf{u} = 0,
\end{equation}
where $\mathbf u$ is the velocity field, $p$ is the pressure, $\mathbf f$ is the volume force arises from the interactions of the suspending fibers, and $Re = \rho\dot{\gamma}L^2/\eta$ is the Reynolds number, where $\rho$ is the density of the fluid, $\eta$ is the dynamic viscosity of the suspending fluid, and $L$ is the characteristic length scale which is also the fiber length.

\subsubsection{Dynamics of flexible slender bodies}

As the fibers are considered continuous one-dimensional objects, Euler-Bernoulli equations can be derived for the motion of flexible fibers as \citep{segel2007mathematics} : 
\begin{equation}
\Delta \rho \frac{{{\partial ^2}{\mathbf{X}}}}{{\partial {t^2}}} = \frac{\partial }{{\partial s}}(T\frac{{\partial {\mathbf{X}}}}{{\partial s}}) - \frac{{{\partial ^2}}}{{\partial {s^2}}}({B^*}\frac{{{\partial ^2}{\mathbf{X}}}}{{\partial {s^2}}}) + \Delta \rho {\mathbf{g}} - {\mathbf{F}} + {{\mathbf{F}}^{f}},
\label{non-dimensional_fil}
\end{equation}
where $s$ is the curvilinear coordinate along the fiber, $\mathbf X$ is the position of the Lagrangian points on the fiber axis, $T$ the tension, $B^* = EI$ the bending rigidity with $E$ the modulus of elasticity of the fiber and $I$ the second moment of inertia around the filament axis, $\mathbf{g}$ is the gravitational acceleration, $\mathbf{F}$ is the fluid-solid interaction force per unit length on the fiber by the surrounding fluid, and  $\mathbf{F}^f$ is the net interaction force on the fiber due to neighboring fibers. Finally, $\Delta \rho$ is the linear density difference between the fiber and the surrounding fluid defined as: 
\begin{equation}
\Delta \rho  = {\rho _f} - \rho {A_f},
\label{density}
\end{equation}
where $\rho_f$ is the actual fiber linear density, and $A_f$ is the sectional area of the fiber. The fibers are considered as inextensible but can bend \citep{huang2007simulation,pinelli2017pelskin}. The inextensible constraint is expressed as : 
\begin{equation}
 \frac{{\partial \mathbf{X}}}{{\partial s}}.\frac{{\partial \mathbf X}}{{\partial s}} = 1.
 \label{inextensible}
\end{equation}
For the case of neutrally buoyant fiber, $\Delta \rho = 0$. Therefore, the left-hand side and the gravitational term on the right-hand side go to zero. For a neutrally buoyant case, the non-dimensional form  of Eq.~\ref{inextensible} remains unchanged and Eq.~\ref{non-dimensional_fil}
can be expressed as : 
\begin{equation}
    0 = \frac{\partial }{{\partial s}}(T\frac{{\partial {\mathbf{X}}}}{{\partial s}}) - \frac{{{\partial ^2}}}{{\partial {s^2}}}({B}\frac{{{\partial ^2}{\mathbf{X}}}}{{\partial {s^2}}}) - {\mathbf{F}} + \mathbf {F}^f
\label{intermediate}    
\end{equation}
with the following characteristics scales: $L$ for length, $U_{\infty} = \dot{\gamma}L$ for velocity, $\dot{\gamma}^{-1}$ for time, $\rho_f$ for reference density, $\rho_f L^2\dot{\gamma}^2$ for tension, and $\rho_f \dot{\gamma}^2L$ for force. Therefore, the dimensionless bending stiffness $B=\frac{B^*}{\rho_f\dot{\gamma}^2L^4}$ measures the ratio of the convective time scale to the elastic time scale. So, as  $B$ decreases, fibers become more flexible. In this study, we fix $B$ at a higher value which ensures a negligible bending of the fiber. Since the left-hand side of Eq.~\ref{intermediate} is zero, in order to avoid singularity in the coefficient matrix, the equation is modified as:  

\begin{equation}
\frac{{{\partial ^2}\mathbf X}}{{\partial {t^2}}} = \frac{{{\partial ^2}{\mathbf X_{fluid}}}}{{\partial {t^2}}} + \frac{\partial }{{\partial s}}(T\frac{{\partial \mathbf X}}{{\partial s}}) - B\frac{{{\partial ^4}\mathbf X}}{{\partial {s^4}}} - \mathbf F + \mathbf {F}^f,
\label{pinal}
\end{equation}
where the first term on the right-hand side is the fluid particle acceleration which is identical to the left-hand side for the neutrally buoyant fibers. As the fibers are suspended in the fluid medium, we impose zero force, moment, and tension at the free ends.
\begin{equation}
    \frac{{{\partial ^2}\mathbf X}}{{\partial {s^2}}} = 0, \frac{{{\partial ^3}\mathbf X}}{{\partial {s^3}}} = 0, T = 0
\label{bc_fiber}    
\end{equation}

Finally, We use the immersed boundary method (IBM) \citep{peskin1972flow} to couple the fluid and solid fibers' motions. For the details of the numerical method, the readers are referred to our previous publication \citep{khan2021rheology}.

\subsection{\label{sec:short}Short range interactions:\protect\\ }
Even though the hydrodynamic interactions are well resolved using the IBM, the short-range interactions need a fine Eulerian mesh that increases computational cost. So, we use the proposed model to calculate the short-range interactions. The short-range interaction, $\mathbf F^f=\mathbf F^{lc} +\mathbf F^c$, is split into lubrication correction $\mathbf{F}^{lc}$ and contact force $\mathbf{F}^c$, respectively. The implementation of lubrication correction $\mathbf{F}^{lc}$ can be found in our previous study \citep{khan2021rheology}.

\subsubsection{Contact force}

\begin{figure}
\centering
\includegraphics[width=1.0\linewidth]{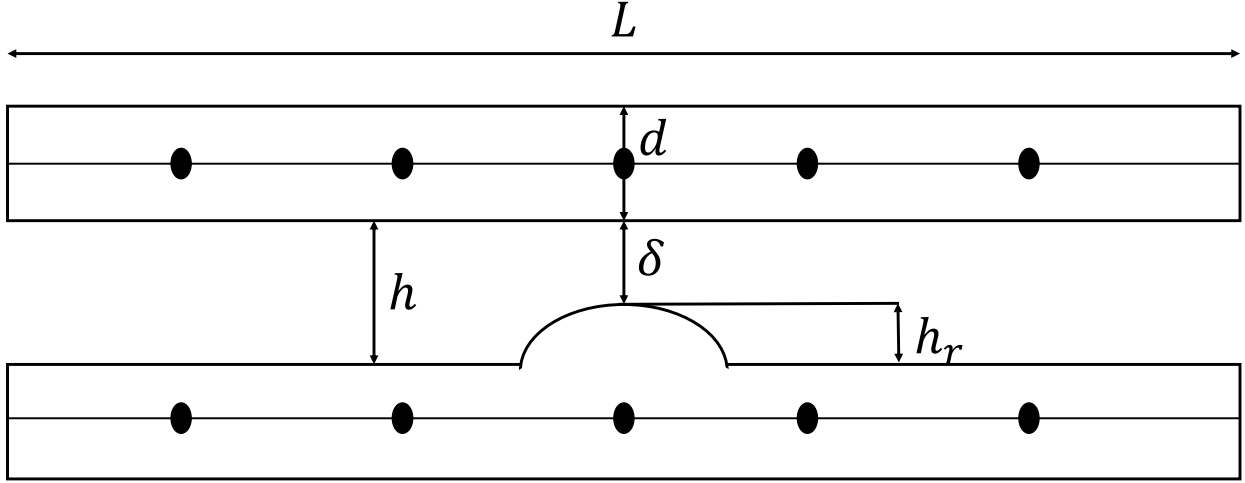}
\caption{A sketch of the roughness model, $L$ and $d$ are the length and diameter of the fiber, respectively, $h_r$ is the roughness height, and $\delta=h-h_r$ is the surface overlap. Contact occurs when $\delta \le 0$. Dots along the axes of the fibers indicate Lagrangian points.}\label{fig:fiber}
\end{figure}

With increasing volume fraction, the surrounding fibers hinder the free rotation of fibers, giving rise to fiber-fiber contacts that influence the microstructure. Microstructure influences the macroscopic rheological properties of the suspension, such as relative viscosity. We model the contact interaction between the fibers as it is done in the discrete element method (DEM). We assume that the contact between the fibers takes place through hemispherical asperity. The asperity deformation is defined by surface overlap $\delta=h-h_r$, and contact happens when $\delta \leq 0$. Here, $h$ is the inter-fiber surface separation, and $ h_r$ is the surface roughness, as shown in figure~\ref{fig:fiber}. 
The deformation of asperities
results in normal, $\mathbf F_n$, and tangential, $\mathbf F_t$ force on the fiber surface. The normal contact force $\mathbf F_n$ is given by the Hertz law:
\begin{equation}\label{eq:hertz}
\mathbf F_n=- F_0\left(\frac{|\delta|}{L}\right)^{3/2} \mathbf{n},
\end{equation}
where $F_0/L^{3/2}$ is the normal stiffness that can be evaluated as a function of fiber's mechanical properties such as elastic modulus, Poisson's ratio, Young's modulus, and roughness size \citep{lobry2019shear, more2020effect}. In this study, we use $F_0$ as the characteristic contact force scale. Coulomb's friction law gives the tangential force:
\begin{equation}
\mathbf F_t=\mu|\mathbf F_n|\frac{\mathbf F_t}{|\mathbf F_t|},
\end{equation}
where $\mu$ is the friction coefficient. 
\subsubsection{Friction model}
\begin{figure}
\centering
\includegraphics[width=1.0\linewidth]{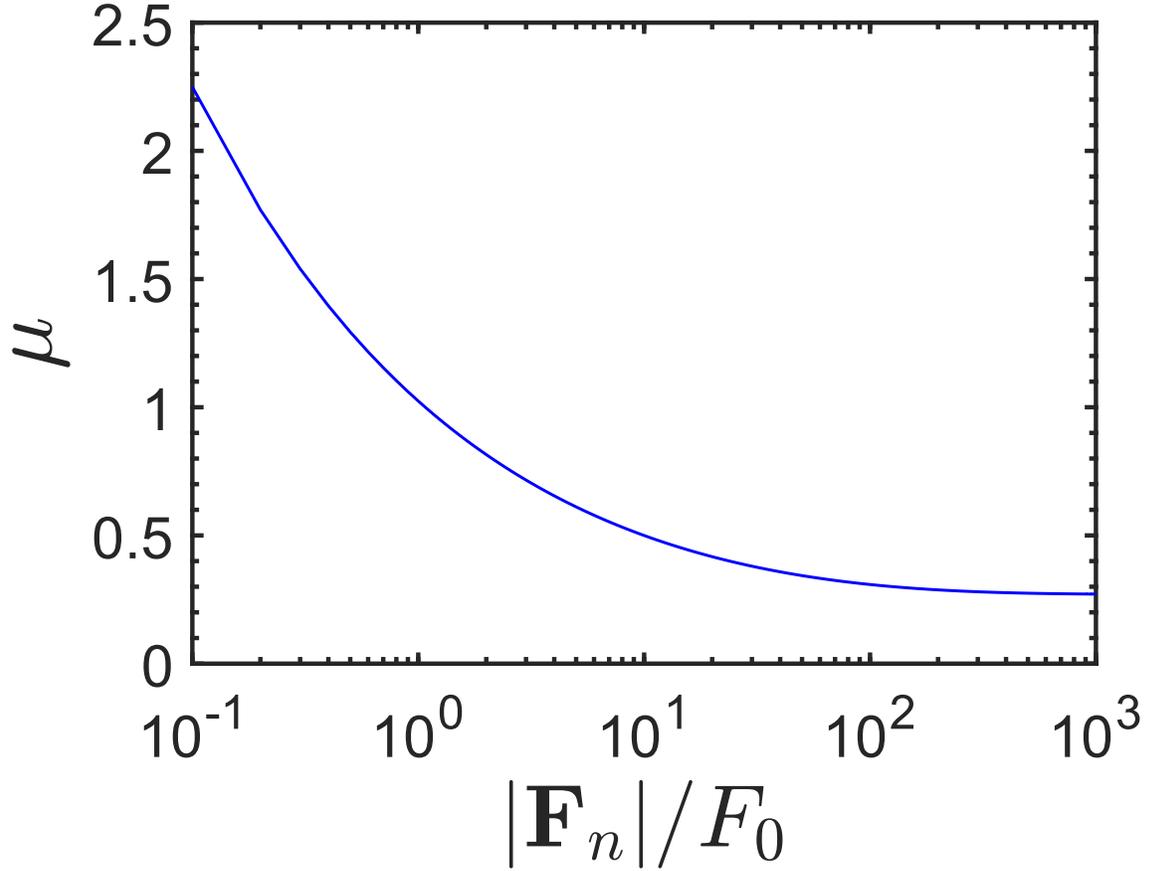}
\caption{Friction coefficient $\mu$ as a function of the dimensionless contact normal force (Eq. \ref{eq:frictionlaw}).}\label{fig:mu_normal_force}
\end{figure}

Researchers have utilized a constant friction coefficient when examining the dynamics of fiber suspensions numerically \citep{stickel2009rheology,banaei2020numerical}. However, a constant coefficient fails to predict the shear rate-dependent suspension viscosity.  In practice, the coefficient of friction depends on many factors, such as the fiber material and the roughness size, which are implicitly included in the normal force via the normal stiffness $F_0/L^{3/2}$ \citep{lobry2019shear, more2020effect}. Hence, a normal load-dependent coefficient of friction is a more accurate physical description of $\mu$ than a simple constant. We use the Brizmer model \citep{brizmer2007elastic} for $\mu$, derived from the measurements between a hemisphere and a flat surface and validated using the finite element analysis \citep{brizmer2007elastic}, which makes it applicable to a wide range of materials and conditions \citep{brizmer2007elastic,lobry2019shear,more2020effect}. It is given by:
\begin{equation}\label{eq:frictionlaw}
\mu  = 0.27\coth \left[ {{0.27{\left({\frac{{|\mathbf F_n^{(i,j)}|}}{{{ F_0}}}} \right)}^{0.35}}} \right],
%\label{eq:variable_friciton}
\end{equation}
where $F_0$ is the characteristic contact force scale introduced in Eq.~\ref{eq:hertz}. Eq.~\ref{eq:frictionlaw} is a decreasing function of $|\mathbf F_n^{(i,j)}|/F_0$ as illustrated in figure~\ref{fig:mu_normal_force}.
Thus, the coefficient of friction decreases with increasing the normal force (which is equivalent to increasing asperity deformation from Eq.~\ref{eq:hertz}) between the contacting fibers and attains a plateau at high normal load values. Before we address the fibers with load-dependent friction coefficient, we will present results with constant $\mu$. This will help to quantitatively understand the original case of load-dependent friction.

\subsection{Stress and bulk rheology calculation}
We compute the bulk stress in the suspension by volume averaging the viscous fluid stress and the stress generated by the presence of fibers and inter-fiber interactions. The calculation of bulk stress, including different contributions, is described elsewhere \citep{banaei2020numerical,khan2021rheology}. In this work, we present the relative viscosity to define the rheological behavior of
the suspension. The relative viscosity $\eta_r$ is defined as: 
\begin{equation}
     \eta_r = \frac{\eta_{eff}}{\eta},
\end{equation}
where $\eta_{eff}$ is the effective viscosity of the suspension. The relative viscosity in terms of bulk stress is: 
\begin{equation}
     \eta_r  = 1 + {\Sigma}{_{xy}^f},
\end{equation}
where ${\Sigma}{_{xy}^f}$ is time and space averaged shear stress arising from the presence of the fibers. ${\Sigma}{_{xy}^f}$ is non-dimensionalized by the product of suspending fluid viscosity $\eta$ and shear rate $\dot\gamma$. As $F_0$ is considered as the characteristic contact force scale, the shear rate scale is given by $\dot{\gamma}_0 = F_0/\pi\eta d^2$.  The dimensionless shear rate provides an estimate of the relative importance of contact to the hydrodynamic forces defined as:
\begin{equation}
    \dot {\Gamma}  = \frac{\dot{\gamma} }{\dot{\gamma}_0} =\frac{\dot{\gamma}}{F_0/\pi \eta d^2} ,
\label{gamma}
\end{equation}
Being the characteristic stress scale  $\sigma_0 = F_0/\pi d^2$, the dimensionless stress can be defined as:
\begin{equation}
    \tilde{\sigma}= \frac{\sigma}{\sigma_0}=\frac{\dot{\gamma} }{\dot{\gamma}_0}{\Sigma}{_{xy}} =\dot {\Gamma}{\eta_r} ,
\label{sigma}
\end{equation}

\subsection{Simulation conditions}
\subsubsection{Boundary conditions and domain size}
\begin{figure}
  \centerline{\includegraphics[width=0.8\linewidth]{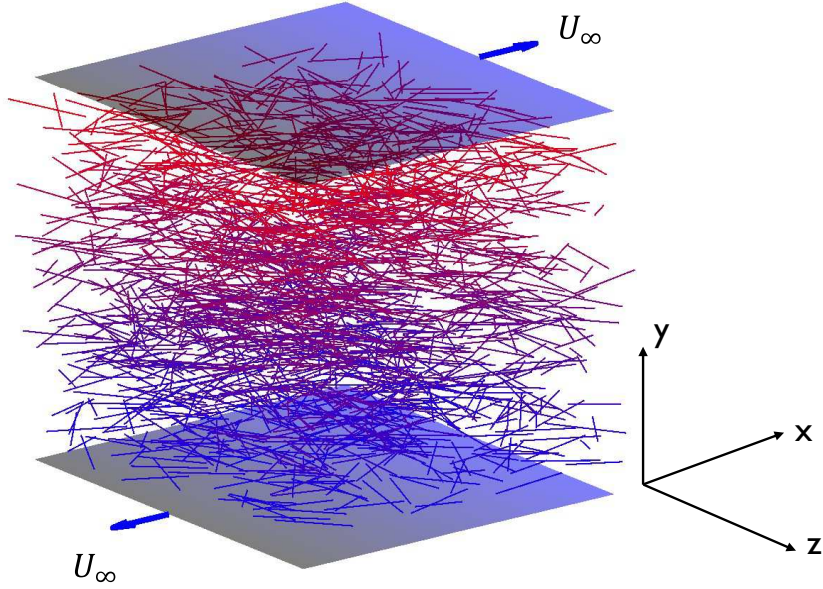}}% Images in 100% size
  \caption{Simulation setup of the shear flow of a fiber suspension. The top and bottom walls move with velocities $U_{\infty} = \dot{\gamma}L$ in the directions shown by the arrows.}
\label{fig:geometry}
\end{figure}

The fibers are suspended in a channel with upper and lower walls moving in the opposite direction with a magnitude of $U_{\infty}=\dot{\gamma}L$ in the x-direction. The wall is subjected to no-slip and no-penetration boundary conditions, and periodicity is assumed in the stream-wise ($x$) and span-wise ($z$) directions. Initially, we place fibers randomly in the simulation domain of size  $5L\times5L\times8L$ and $80\times80\times128$ grid points in the stream-wise ($x$), wall normal ($y$), and span-wise direction ($z$), respectively. A schematic diagram of the computational configuration and coordinate system is shown in  figure~\ref{fig:geometry}. The averaged steady state suspension viscosity changes minimal (less than 2 \%) in simulations with bigger domain,  higher grid, and time resolutions such as 1.5, 2, 2.5, and 3 times the current domain, grid, and time resolution, Moreover, 36 Lagrangian points over the fiber length are enough to resolve the case with the highest fiber aspect ratio. The required time step to capture the fiber dynamics is $\Delta t = 10^{-5}$. The suspension is simulated until a statistically steady viscosity is observed and mean values after discarding the initial transients are presented. 

\subsubsection{Range of parameters investigated} 
The aim of this study is to quantify the effect of varying fiber aspect ratios, volume fractions, and shear rates on the rheology  of the suspension. Hence, we simulate suspensions of almost rigid fibers in a shear flow by varying the aspect ratio in the range of $10 \leq AR \leq 36$. The simulations were carried out for dimensionless shear rate in the range of $0.1\leq \dot\gamma/\dot\gamma_0\leq100$ for a range of volume fractions $0.03 \leq \phi \leq 0.45$. The dimensionless bending rigidity, $B$, was set to 5.0 which ensures a negligible bending of the fiber.
The range of parameters explored in the present work is summarized in table \ref{tab:my-table}.

\begin{table}
\caption{Range of parameters explored in this study}
  %\begin{center}
  \begin{ruledtabular}
  \begin{tabular}{cccc}
      $\phi$  & $AR$   &   $\dot\gamma/\dot\gamma_0$ & $\mu$\\\hline\\
      
       $0.03-0.47$   &~ $10-36$~ & ~$0.1-100$~ &~~ $0-11$~ \\
  \end{tabular}
  %\caption{Range of parameters explored in this study}
  \label{tab:my-table}
\end{ruledtabular}
\end{table}

  %\begin{tabular}{cccccccc}

     % \hline
   %   $10$  &~ $0.40$~ & ~$0.47$~ &~$0.11$~ &~$0.85$~ &~$20.9$~ &~$-7.532$~\\
      %\hline
     % $36$  &~ $0.17$~ & ~$0.12$~ &~$0.11$~ %&~$0.8015$~ &~$0.82$~&~$-7.532$~\\

  %\end{tabular}
  
 % \label{tab:fitting_phi_a}
 % \end{ruledtabular}
%\end{table}

\section{\label{sec:model}The constitutive model:\protect\\ }

We first demonstrate simulations where the friction coefficient is maintained constant before addressing the fibers with a load-dependent friction coefficient. Thus, we reflect on the significant impact of this microscopic parameter on the suspension viscosity. Recent numerical investigations on frictional non-Brownian particle suspensions have shown that the friction coefficient, considered constant, has a significant impact on the effective viscosity \citep{mari2014shear,gallier2014rheology,gallier2018simulations,singh2018constitutive}. From these studies, it is clear that the jamming volume fraction depends significantly on the friction coefficient. More recently, \cite{singh2018constitutive} explained how the viscosity of non-colloidal suspensions changes with the microscopic particle friction coefficient by showing that the jamming volume fraction decreases when the microscopic particle friction coefficient goes from 0 to 1. We point out that simulations of granular flows \citep{silbert2010jamming} and experiments with discontinuous shear-thickening suspensions \citep{fernandez2013microscopic} have shown that there is a link between the amount of jamming and the microscopic friction coefficient. Therefore, we start our analysis by deriving correlation laws for the dependence of suspension jamming fraction with constant friction coefficients. After that, the correlation laws will be utilized to describe the rheology of suspensions for cases with load-dependent friction.

\begin{figure*}
\centering
\begin{subfigure}{.5\textwidth}
  \centering
  \includegraphics[width=1.0\linewidth]{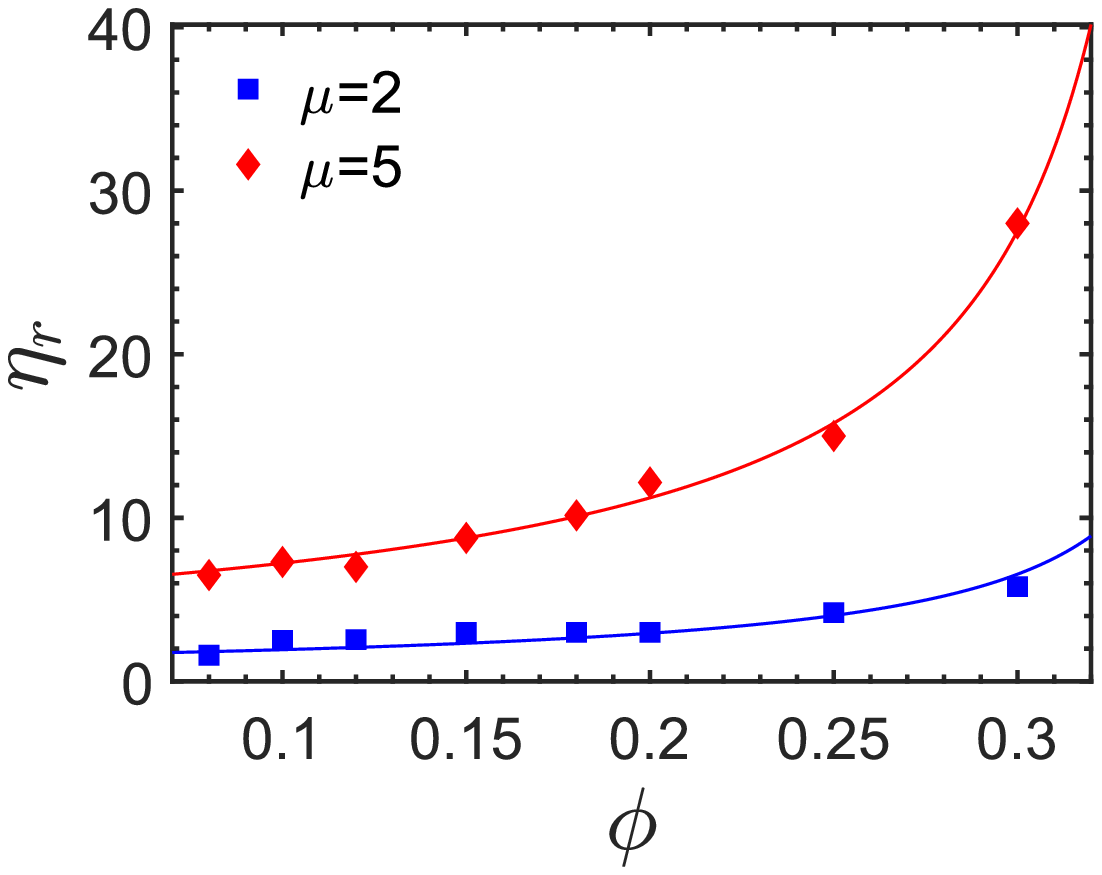}
  \caption{}
  \label{fig:viscosity_vol_mu_const_AR18}
\end{subfigure}%zz
\begin{subfigure}{.5\textwidth}
  \centering
  \includegraphics[width=1.0\linewidth]{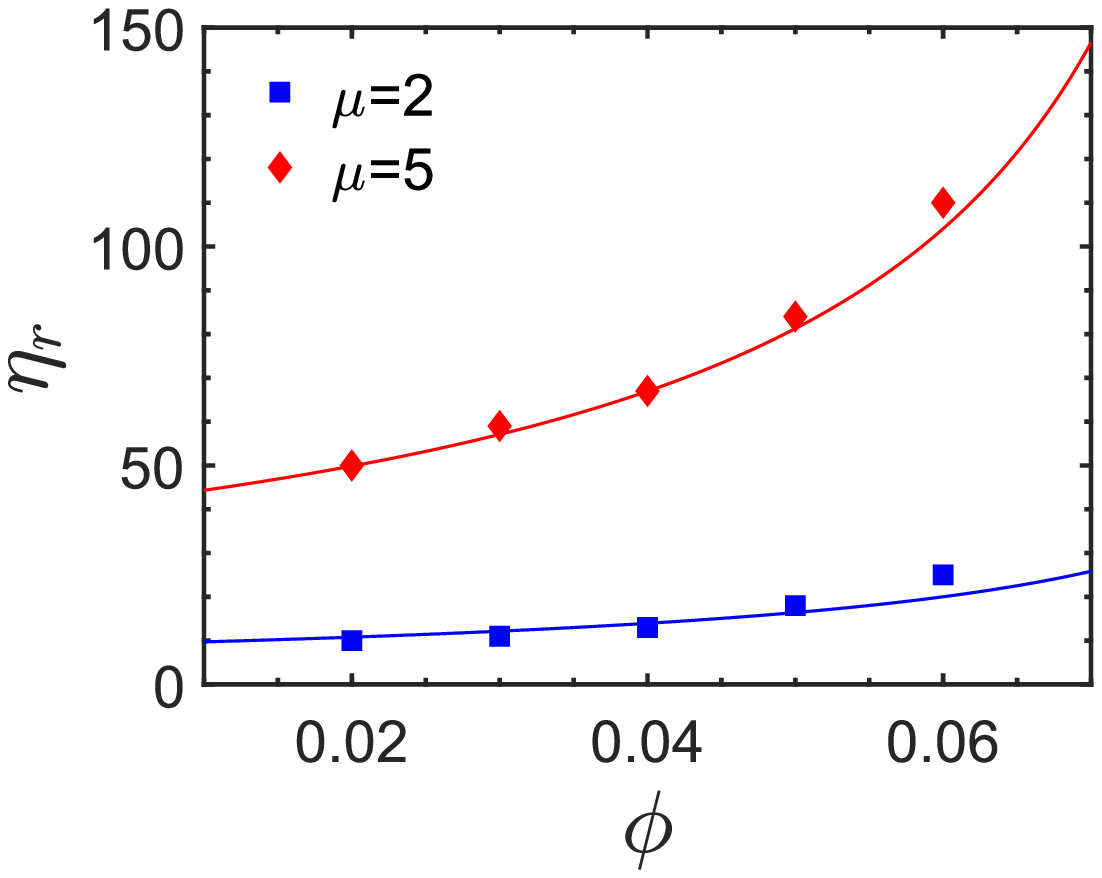}
  \caption{}
  \label{fig:viscosity_vol_mu_const_AR33}
\end{subfigure}
\caption{ Relative viscosity as a function of volume fraction for friction coefficient $\mu =2, 5$ for (a) $AR = 10$ and (b) $AR = 36$. Solid lines: best fit with Eq.~\ref{eq:viscosity_vol}}.
\label{fig:viscosity_vol_mu_const_AR}
\end{figure*}

\begin{figure*}
\centering
\begin{subfigure}{.5\textwidth}
  \centering
  \includegraphics[width=1.0\linewidth]{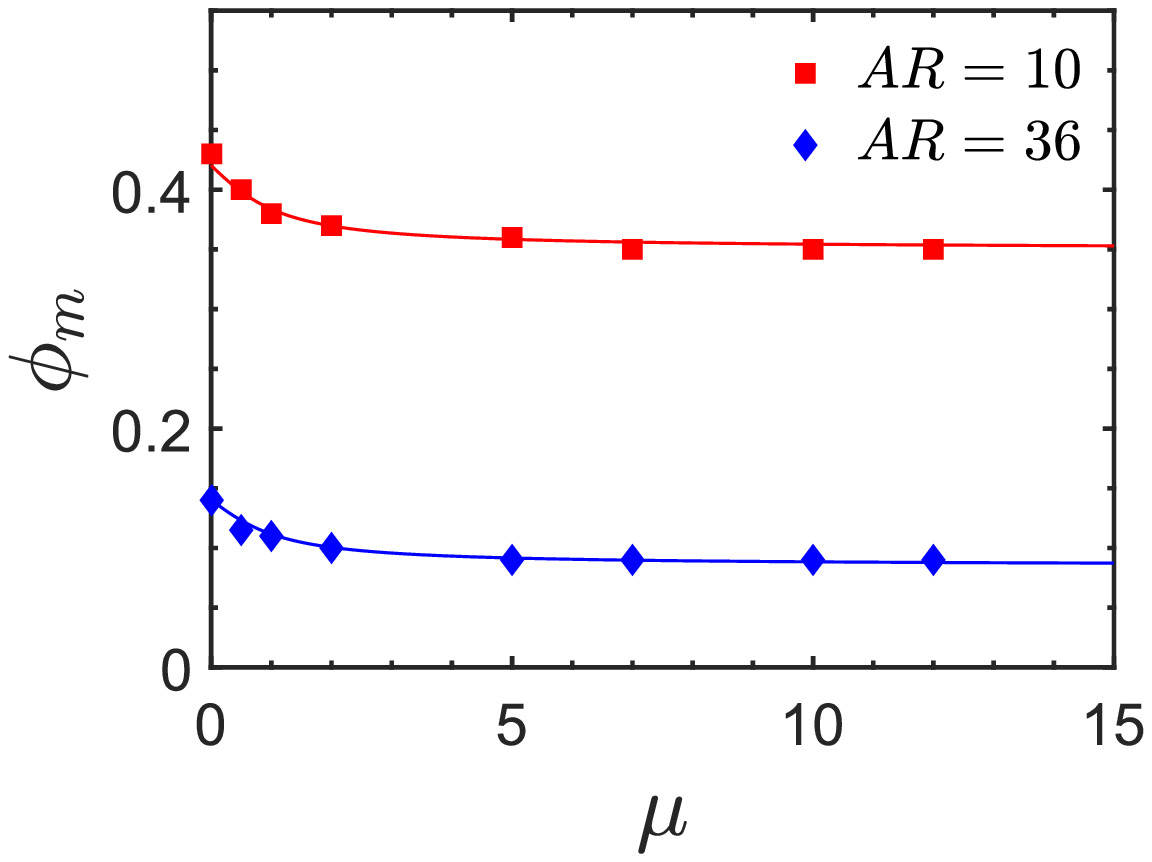}
  \caption{}
  \label{fig:phi_mu}
\end{subfigure}%zz
\begin{subfigure}{.5\textwidth}
  \centering
  \includegraphics[width=1.0\linewidth]{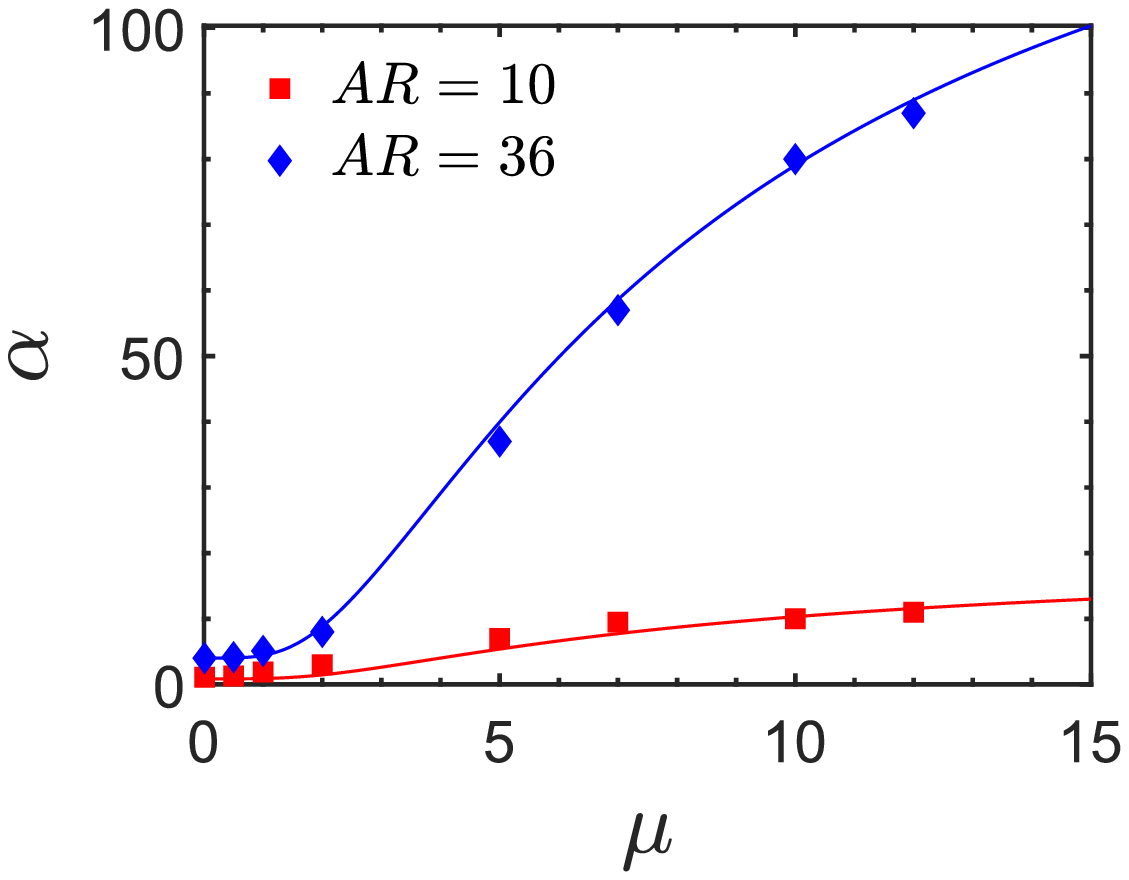}
  \caption{}
  \label{fig:alpha_mu}
\end{subfigure}
\caption{ (a) Jamming volume fraction $\phi_m$ (b) pre-factor $\alpha$ as a function of $\mu$. Solid lines: best fit with Eq.~\ref{eq:constant_fric_mu_alpha}.}.
\label{fig:phi_m_a_mu}
\end{figure*}

In the beginning, we report the simulations for suspensions of fibers in which the friction coefficient is held at a fixed value in the range of 0 to 15. The following correlation law is fitted for each value of $\mu$ to describe the variation of the relative viscosity against the volume fraction.
\begin{equation}
    \eta_r (\mu,\phi,AR) = \alpha(\mu,AR)\bigg(1-\frac{\phi}{\phi_m(\mu, AR)}\bigg)^{-0.90},
    \label{eq:viscosity_vol}  
\end{equation}
where $\alpha$ and $\phi_m$ are friction and aspect ratio dependent constant coefficients. Here, we model the dependence of viscosity on volume fraction as $\phi_m^{-0.90}(\phi_m-\phi)^{-0.90}$. The form is consistent with the proposed relationships in the recent work concerning the suspension of fibers \citep{khan2021rheology,tapia2017rheology}. Figure~\ref{fig:viscosity_vol_mu_const_AR} presents the results of the simulation along with the relevant correlation laws for two different values of the friction coefficient ($
\mu =2$ and $\mu = 5$) for aspect ratios $AR = 10$ and $AR = 36$.
The friction and aspect ratio dependent parameters ($\alpha$ and $\phi_m$) are found empirically from our simulations as:

\begin{eqnarray}
{\alpha(\mu,AR)} =&& \alpha^{\mu_\infty}(AR)+\bigg(\alpha^{\mu_0}(AR)-\alpha^{\mu_\infty}(AR)\bigg)\nonumber\\
&&\frac{\textrm{exp}\big(-X^\alpha \textrm{atan}(\mu)\big)-\textrm{exp}(-\pi X^\alpha /2)}{1-\textrm{exp}(-\pi X^\alpha /2)},
\\
{\phi_m(\mu,AR)} = && \phi_m^{\mu_\infty}(AR)+\bigg({\phi_m}^{\mu_0}(AR)-\phi_m^{\mu_\infty}(AR)\bigg)\nonumber\\
&&\frac{\textrm{exp}\big(-X^q \textrm{atan}(\mu)\big)-\textrm{exp}(-\pi X^q /2)}{1-\textrm{exp}(-\pi X^q /2)}
\label{eq:constant_fric_mu_alpha}
\end{eqnarray}
as shown by fits in figure~\ref{fig:phi_m_a_mu}. Here, the superscript $\mu_0$ and $\mu_\infty$ denote the corresponding parameter when the coefficient of friction is zero and infinity \citep{lobry2019shear}. Note that all the parameters are independent of the coefficient of friction and are reported in table~\ref{tab:fitting_phi_a}.
\begin{table}
\caption{Calibrated model parameters for Eq.~\ref{eq:constant_fric_mu_alpha} for  aspect ratios $AR = 10$ and  $AR = 36$}
  \begin{ruledtabular}
  \begin{tabular}{cccccccc}
  
      $AR$  & $\phi_m^0$   &   $\phi_m^{\infty}$  & $X^q$ &   $\alpha^0$  &$\alpha^{\infty}$  &$X^{\alpha}$\\[1pt]
      \hline
      $10$  &~ $0.40$~ & ~$0.47$~ &~$0.11$~ &~$0.85$~ &~$20.9$~ &~$-7.532$~\\
      %\hline
      $36$  &~ $0.12$~ & ~$0.17$~ &~$0.11$~ &~$0.8015$~ &~$0.82$~&~$-7.532$~\\

  \end{tabular}
  
  \label{tab:fitting_phi_a}
  \end{ruledtabular}
\end{table}
In particular, for aspect ratio 18, the jamming volume fraction seems to approach the limit $\phi_m^{\infty} = 0.47$ as the friction coefficient goes to $\infty$ and is equal to $\phi_m^{0} = 0.40$ for $\mu =0$. These values are consistent with the data found in the literature {\citep{williams2003random, khan2021rheology}}.

\definecolor{red}{rgb}{1,0,0}
\definecolor{blue}{rgb}{0,0,1}

\newcommand{\bt}{\textcolor{blue}{$\blacktriangle$}} 
\newcommand{\bs}{\textcolor{blue}{$\blacksquare$}}

\newcommand{\bc}[2][blue,fill=blue]{\tikz[baseline=-0.5ex]\draw[#1,radius=#2] (0,0) circle ;}%

\newcommand{\rc}{{\color{red}$\circ$}}
\newcommand{\rt}{{\color{red}$\triangle$}}
\newcommand{\rs}{{\color{red}$\square$}}

\begin{figure}
\centering
\includegraphics[width=1.0\linewidth]{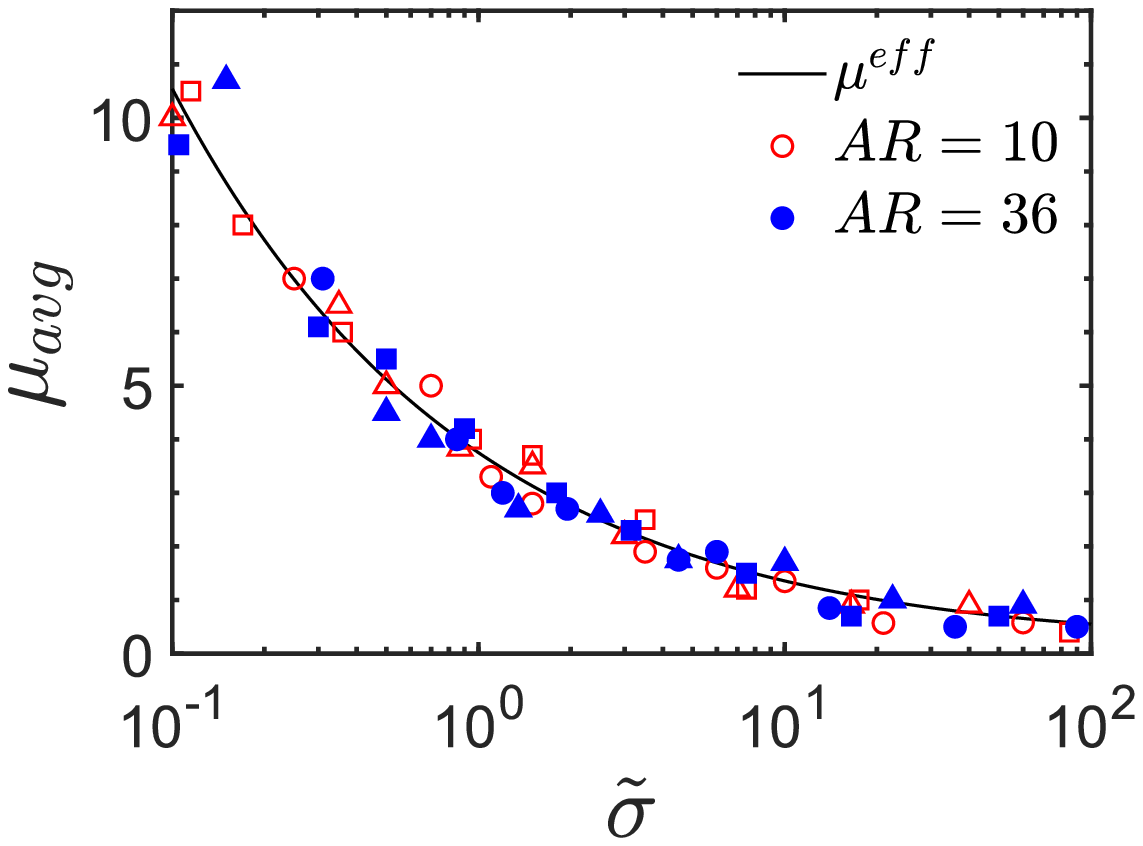}
\caption{Average coefficient of friction over all contacting fiber pairs as a function of shear stress. Red symbol corresponds to data for $AR = 10$ having volume fractions (\rs) $\phi =0.20$, (\rt) $ \phi = 0.30$, and (\rc) $ \phi = 0.35$. Blue symbol corresponds to data for $AR = 36$ having volume fractions (\bs) $\phi =0.03$, (\bt) $\phi = 0.05$, and (\bc{3.0pt}) $\phi = 0.08$. The solid line shows the effective coefficient of friction from Eq.~\ref{eq:mu_stress}. }\label{fig:mu_stress}
\end{figure}

The above analysis shows that the simulations with a constant friction coefficient do not provide rate-dependent viscosity in the suspension. Hence, We now turn to the analysis with load-dependent friction between the fibers in the suspension. However, the correlation derived earlier with a constant coefficient of friction will aid quantitative analysis of the case with a load-dependent friction coefficient.  %Hence it will be useful in pinpointing the exact dependence of the jamming volume fraction on the coefficient of friction

From the shear rate-dependent viscosity, we observe that  as the  reduced  shear rate  increases,  the  viscosity  decreases due to the reduction of friction coefficient \citep{khan2021rheology}. It is possible to gain a better understanding of this transition from high to low viscosity by examining the average friction coefficient, $\mu_{avg}$, as a function of the reduced shear stress, $\tilde{\sigma} $ ($\eta_r\dot{\gamma}/\dot{\gamma_0}$),  as shown in figure~\ref{fig:mu_stress}. Reduced shear stress $\tilde{\sigma} $ quantifies the typical force experienced by  two  contacting  fibers, and $\mu_{avg}$ is the ensemble average of the coefficient of friction $\mu$ between all the contacting fiber pairs in the suspension. Intriguingly, regardless of the fiber volume fraction and aspect ratio, the data collapse onto a single curve, denoted by $\mu^{eff}$, fitted with:
\begin{equation}
    \mu^{eff} = 0.34 \textrm{coth}\Bigg[0.35\bigg(\frac{\tilde{\sigma}}{20}\bigg)^{0.45}\Bigg].
    \label{eq:mu_stress}
\end{equation}
The functional form is similar to the one for the coefficient of friction $\mu$ dependence on the normal load between the fibers $|\mathbf{F}^{(i,k)}_n|$ in Eq.~\ref{eq:frictionlaw}, however, it is not exactly the same. This feature is not immediately apparent; thus, it is worth commenting on. The normal contact force does regulate the microscopic coefficient of friction between contacting fibers. However, the bulk shear stress and mean force are only qualitatively related. Even the relationship between the mean friction coefficient and mean force is difficult to predict due to the highly non-linear relationship between the microscopic friction coefficient and force. This implies that the reduced shear stress may be interpreted as the reduced effective normal force between contacting fibers, which regulates the mean friction coefficient. Therefore, we obtain the following expression for suspension viscosity at a finite stress value in terms of the volume fraction $\phi$, jamming volume fraction $ \phi_m$, and pre-factor $\alpha$.

%(\tilde{\sigma},AR)$ and fitting constant $[\alpha(\tilde{\sigma},AR)$] as

\begin{eqnarray}
    {\eta_r (\tilde{\sigma},\phi,AR)} =&& \alpha\Big(\tilde{\sigma}(\mu_eff),AR\Big)\nonumber\\
    &&\bigg(1-\frac{\phi}{\phi_m\Big(\tilde{\sigma}(\mu_eff)), AR\Big)}\bigg)^{-0.90}
    \label{eq:viscosity}  
\end{eqnarray}
The rheological properties in the two extreme stress conditions can be expressed in terms of volume fractions $\phi$, jamming volume fractions  $\phi_m^{0,\infty}$ and model parameters $\alpha^{0,\infty}$. Here the superscript 0 and $\infty$ denote the low and high shear limits, respectively.
\begin{eqnarray}
\eta_r^0 (\phi,AR) = \alpha^0(AR)\bigg(1-\frac{\phi}{\phi_m^0(AR)}\bigg)^{-0.90},
\label{eq:low_shear}
\\
\eta_r^{\infty}(\phi,AR) = \alpha^{\infty}(AR)\bigg(1-\frac{\phi}{\phi_m^{\infty}(AR)}\bigg)^{-0.90}
\label{eq:high_shear}
\end{eqnarray}
The jamming fraction $\phi_m$ and the pre-factor $\alpha$ at intermediate stress $\tilde{\sigma}$ can be calculated by interpolating their corresponding values in the low and high-stress limits \citep{more2020constitutive} as follows: 

\begin{eqnarray}
 \phi_m(\sigma,AR) = \phi_m^0(AR)[1-f
 (\tilde{\sigma})]+\phi_m^{\infty}
 [f(\tilde{\sigma})] 
 \label{eq:jamming_phi}
\\
 \alpha(\sigma,AR) = \alpha^{\infty}(AR)  [f_1(\tilde{\sigma})] +\alpha^0[1-f_1
 (\tilde{\sigma})],
 \end{eqnarray}
where 
\begin{eqnarray}
f(\tilde{\sigma})&=& f\Big(\tilde{\sigma}(\mu_{eff})\Big)\nonumber\\ =&&\frac{\textrm{exp}\big(-X^q \textrm{atan}(\tilde{\sigma}(\mu_{eff}))\big)-\textrm{exp}(-\pi X^q /2)}{1-\textrm{exp}(-\pi X^q /2)}, 
\\
f_1(\tilde{\sigma}) &=& f\Big(\tilde{\sigma}(\mu_{eff})\Big) \nonumber\\ =&& \frac{\textrm{exp}\big(-X^\alpha \textrm{atan}(\tilde{\sigma}(\mu_{eff}))\big)-\textrm{exp}(-\pi X^{\alpha} /2)}{1-\textrm{exp}(-\pi X^{\alpha} /2)} .
 \end{eqnarray}

%Here, $f(\tilde{\sigma}) = f\Big(\tilde{\sigma}(\mu_{eff})\Big) = \frac{\textrm{exp}\big(-X^q \textrm{atan}(\tilde{\sigma}(\mu_{eff}))\big)-\textrm{exp}(-\pi X^q /2)}{1-\textrm{exp}(-\pi X^q /2)} $, and $f_1(\tilde{\sigma}) = f\Big(\tilde{\sigma}(\mu_{eff})\Big) = \frac{exp\big(-X^q \textrm{atan}(\tilde{\sigma}(\mu_{eff}))\big)-\textrm{exp}(-\pi X^{\alpha} /2)}{1-\textrm{exp}(-\pi X^{\alpha} /2)} $. %The value of $X^{\alpha,q}$ has been been mentioned in table~\ref{tab:fitting_phi_a}. 
A similar interpolation function was used to interpolate $\alpha$ and $\phi_m$ when we considered results with a constant coefficient of friction. Later we find an expression to describe the reduced stress as a function of $u^{eff}$ (Eq.~\ref{eq:mu_stress}). Thus, we use a similar interpolation function as expressed by $f(\tilde{\sigma})$. In addition, the pre-factor $\alpha$ and jamming volume fraction $\phi_m$ in the low and high shear stress limits can be expressed in terms of aspect ratio, $AR$ and model parameters \{$\bar{\phi}_m, \bar{\alpha}\}^{0,\infty}$, \{$\hat{\phi}_m, \hat{\alpha\}}^{0,\infty}$,  \{${Q}_{\alpha,\phi_m}\}^{0,\infty}$, \{${R}_{\alpha,\phi_m}\}^{0,\infty}$, and \{${S}_{\alpha,\phi_m}\}^{0,\infty}$. Here, the superscript 0 and $\infty$ denote the model parameters at low and high shear limits, respectively. We use $\bar{}$ and $\hat{}$ over model parameters for the fibers with aspect ratios 10 and 36 cases, respectively. 

\begin{eqnarray}
{\alpha^{0,\infty}}=&&\bar{\alpha}^{0,\infty}+\Big[\hat{\alpha}^{0,\infty} - \bar{\alpha}^{0,\infty}\Big]\nonumber\\
&&\textrm{log}\Big(\frac{Q_{\alpha}^{0, \infty}}{{\left( AR\right)}^{R_{\alpha}^{0,\infty}}}\Big)^{S_{\alpha}^{0,\infty}}
\label{eq:alpha_aspect}
\\
{\phi_m^{0,\infty}}=&&\bar{\phi_m}^{0,\infty}+\Big[\hat{\phi_m}^{0,\infty} - \bar{\phi_m}^{0,\infty}\Big]\nonumber\\
&&\textrm{log}\Big(\frac{Q_{\phi_m}^{0, \infty}}{{\left( AR\right)}^{R_{\phi_m}^{0,\infty}}}\Big)^{S_{\phi_m}^{0,\infty}}
\label{eq:phi_aspect}
\end{eqnarray}
The value of the calibrated model parameters is reported in table~\ref{tab:alpha_phi_aspect}.
\section{\label{sec:result}result}

\subsection{Aspect ratio dependent rheology}
\begin{figure}
\centering
\includegraphics[width=1.0\linewidth]{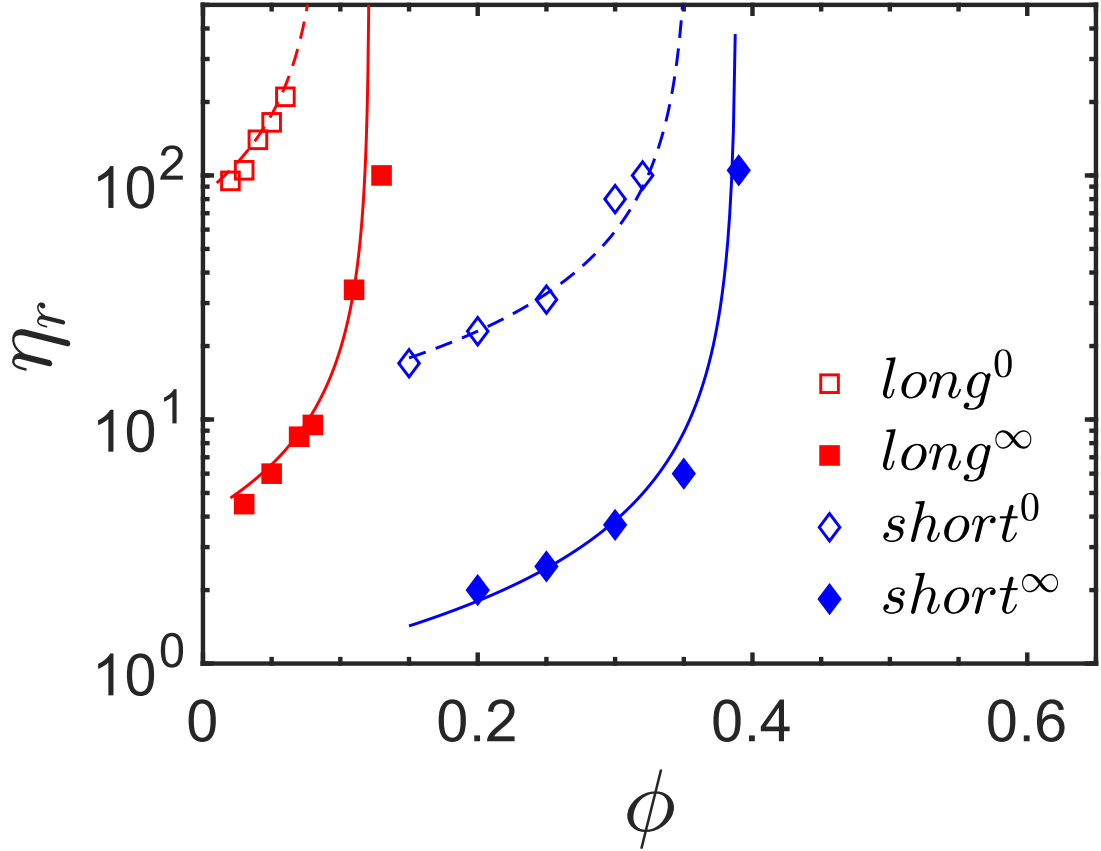}
\caption{ Relative viscosity of the long ($AR = 36$) and short ($AR = 10$) fiber suspensions for different volume fractions. Dashed and solid lines represent fitting Eqs. in the low $(^0$, Eq.~\ref{eq:low_shear}$)$ and high $(^{\infty}$, Eq.~\ref{eq:high_shear}$)$ stress limits. }\label{fig:jamming}
\end{figure}

\begin{figure*}
\centering
\begin{subfigure}{.5\textwidth}
  \centering
  \includegraphics[width=1.0\linewidth]{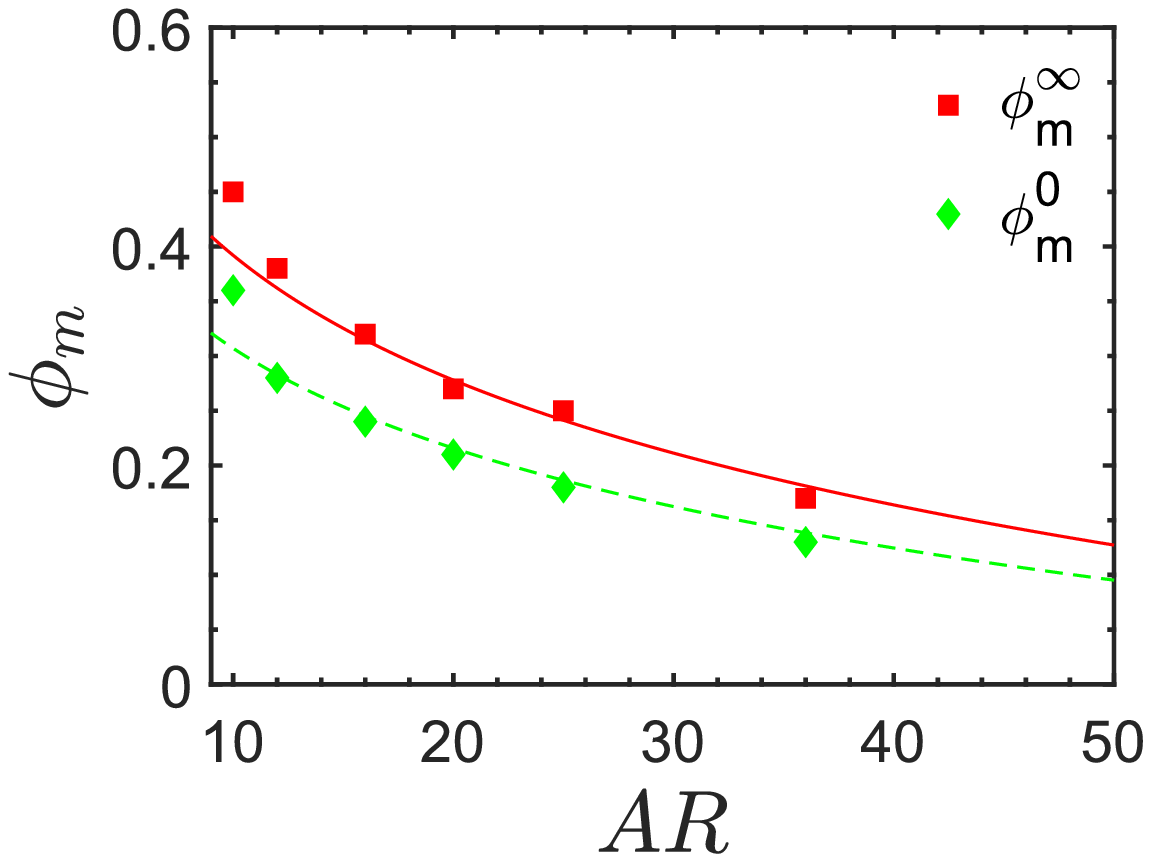}
  \caption{}
  \label{fig:phi_AR}
\end{subfigure}%zz
\begin{subfigure}{.5\textwidth}
  \centering
  \includegraphics[width=1.0\linewidth]{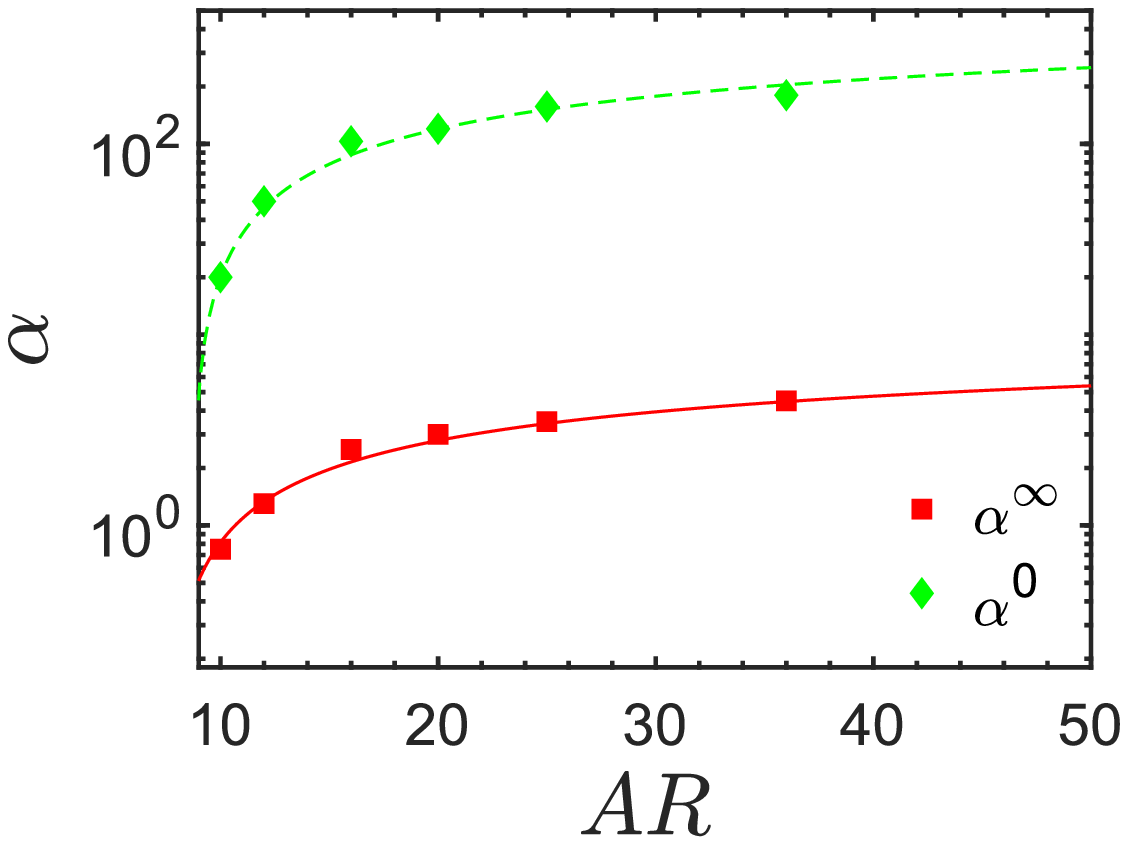}
  \caption{}
  \label{fig:alpha_AR}
\end{subfigure}
\caption{ Calibrated model parameters as a function of the aspect ratio of the fibers. Dashed and solid lines represent fitting Eqs. (\ref{eq:alpha_aspect}-\ref{eq:phi_aspect}) in the low $(^0)$ and high $(^{\infty})$ shear limits. (a) jamming volume fraction, $\phi_m$ and (b) calibrated model parameter, $\alpha$. An increase in the aspect ratio leads to an increase in $\alpha$ in both the low and high shear rate limits. However, increasing the aspect ratio reduces the jamming volume fraction, $\phi_m$ due to the increase in relative viscosity \citep{keshtkar2009rheological, tapia2017rheology, khan2021rheology} .}
\label{fig:phi_alpha_AR}
\end{figure*}

Experiments \citep{keshtkar2009rheological, bounoua2016shear} and numerical simulations \citep{khan2021rheology,tapia2017rheology} have shown that increasing aspect ratio leads to increased suspension viscosity due to a reduction of the jamming fraction. Figure~\ref{fig:jamming} shows the relative viscosity as a function of fiber volume fraction for the long ($AR = 36$) and short ($AR = 10$) fibers in the low and high shear rate limits with the modified Maron-Pierce fitting curves [Eqs.~\ref{eq:low_shear},\ref{eq:high_shear}]. As we present for $AR = 10$ and $AR = 36$ in the main text, we refer to the cases as "short" and "long" henceforth in the article. Here, the reduction in the jamming volume fraction, $\phi_m$, with increasing aspect ratio and stress is consistent with experiments \citep{keshtkar2009rheological, bounoua2016apparent, bounoua2016shear,  tapia2017rheology}. We observe that the calibrated model parameters and jamming volume fraction depend only on the fiber aspect ratio in the low and high shear rate limits. So, they can be expressed in terms of $AR$, as shown in Eqs.~(\ref{eq:alpha_aspect})-(\ref{eq:phi_aspect}). Figures~\ref{fig:phi_AR} and  \ref{fig:alpha_AR} show that Eqs.~(\ref{eq:alpha_aspect})-(\ref{eq:phi_aspect}) are a good fit and accurately capture the effect of increasing fiber aspect ratio on the rheology of the dense fiber suspensions in the low and high shear limits.

\begin{table*}
\caption{Aspect ratio dependent calibrated model parameters for relative viscosity, $\eta_r$. }
  \begin{ruledtabular}
\def~{\hphantom{1}}
  \begin{tabular}{ccccccccccc}
 
      ${}$  & ${\{ \bar{} \}}^0$   &   ${\{\hat{}\}}^0$  & $Q_{\{\}}^0$ &   $R_{\{\}}^0$  &$S_{\{\}}^0$  &${\{ \bar{} \}}^{\infty}$   &   ${\{\hat{}\}}^{\infty}$  & $Q_{\{\}}^{\infty}$ &   $R_{\{\}}^{\infty}$  &$S_{\{\}}^{\infty}$\\[1pt]
      \hline\\
      $\alpha$  &~ $145$~ & ~$15$~ &~$2.04$~ &~$1.3$~ &~$0.85$~ &~$4$~&~$0.75$~&~$82.35$~&~$1.25$~&~$0.70$~\\
     % \hline
      $\phi_m$  &~ $0.23$~ & ~$0.70$~ &~$57$~ &~$1.4$~ &~$0.20$~&~$0.2$~&~$0.57$~&~$82$~&~$1.27$~&~$0.35$~\\

  \end{tabular}
  
  \label{tab:alpha_phi_aspect}
   \end{ruledtabular}
\end{table*}

Before going into detail about stress-dependent rheology, it is important to note that aspect ratio dependence and stress dependence are unrelated. Stress-dependent rheological behavior is recovered using a load-dependent friction coefficient model, whereas aspect ratio dependence is obtained by altering the $AR$ of the fiber. In the absence of load-dependent friction, we only get the aspect ratio dependence rheology as presented in figure~\ref{fig:jamming} and modeled in Eqs.~(\ref{eq:alpha_aspect})-(\ref{eq:phi_aspect}). We will get stress between 0 and $\infty$ depending on the value of $\mu$. But the increase in viscosity with aspect ratio will still be observed, which is consistent with earlier simulations \citep{wu2010numerical, khan2021rheology} and experiments \citep{keshtkar2009rheological,bounoua2016shear, tapia2017rheology}.

\subsection{Stress dependent viscosity}

In this section, we present the shear stress-dependent viscosity for suspensions with varying fiber aspect ratios along with the constitutive equations fitting curves. It has been demonstrated that the rheological properties in the intermediate stress values can be interpolated once the rheological characteristics and the jamming fraction in the low and high shear stress limits are known \citep{singh2018constitutive,more2020constitutive}. Section \ref{sec:model} proposes aspect ratio and stress-dependent constitutive equations based on this premise.

\begin{figure*}
\centering
\begin{subfigure}{.5\textwidth}
  \centering
  \includegraphics[width=1.0\linewidth]{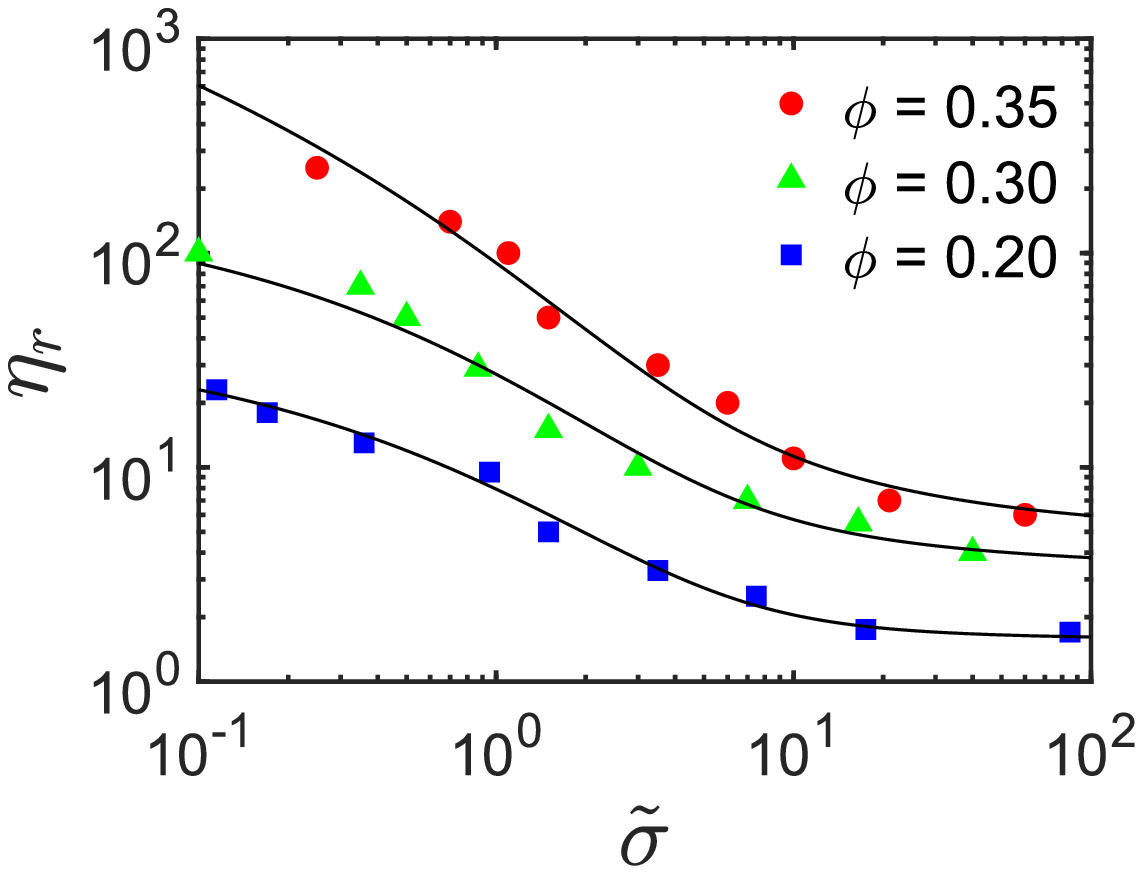}
  \caption{}
  \label{fig:viscosity_stress_AR_10}
\end{subfigure}%zz
\begin{subfigure}{.5\textwidth}
  \centering
  \includegraphics[width=1.0\linewidth]{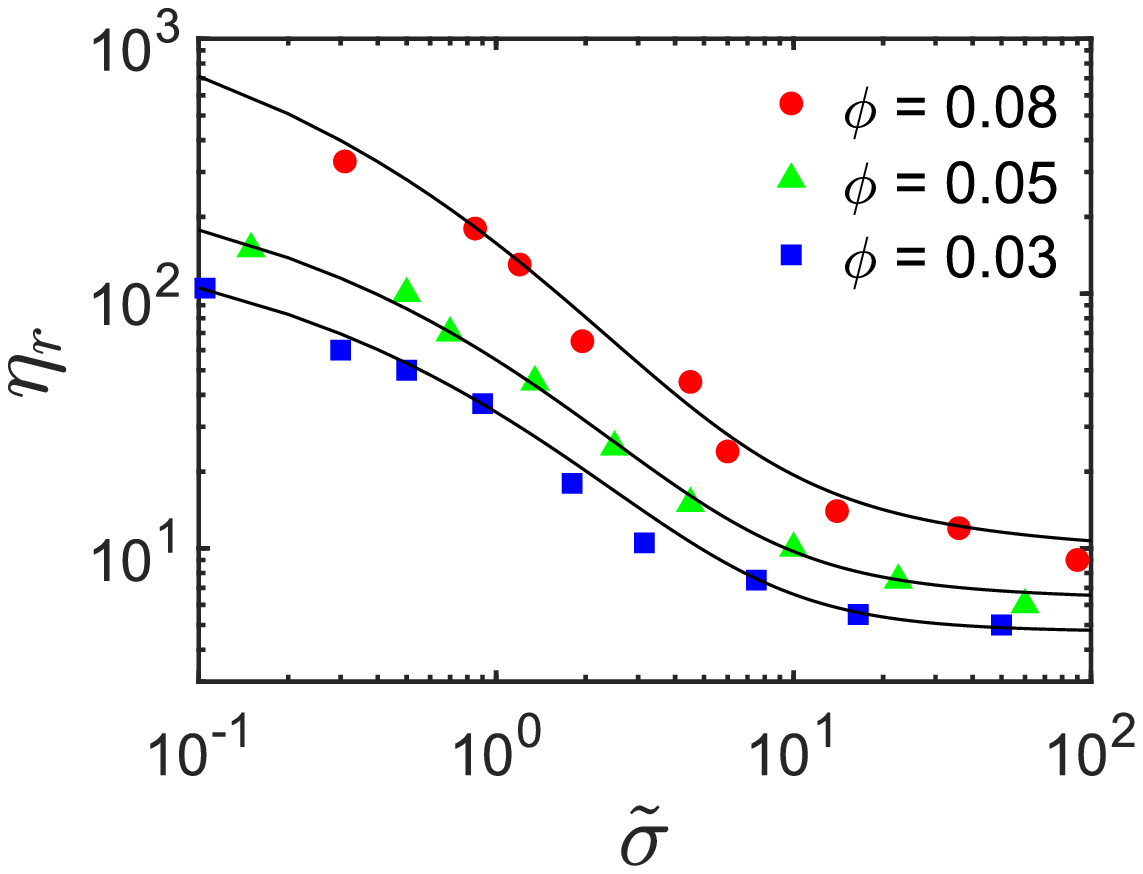}
  \caption{}
  \label{fig:viscosity_stress_AR_36}
\end{subfigure}
\begin{subfigure}{.5\textwidth}
  \centering
  \includegraphics[width=1.0\linewidth]{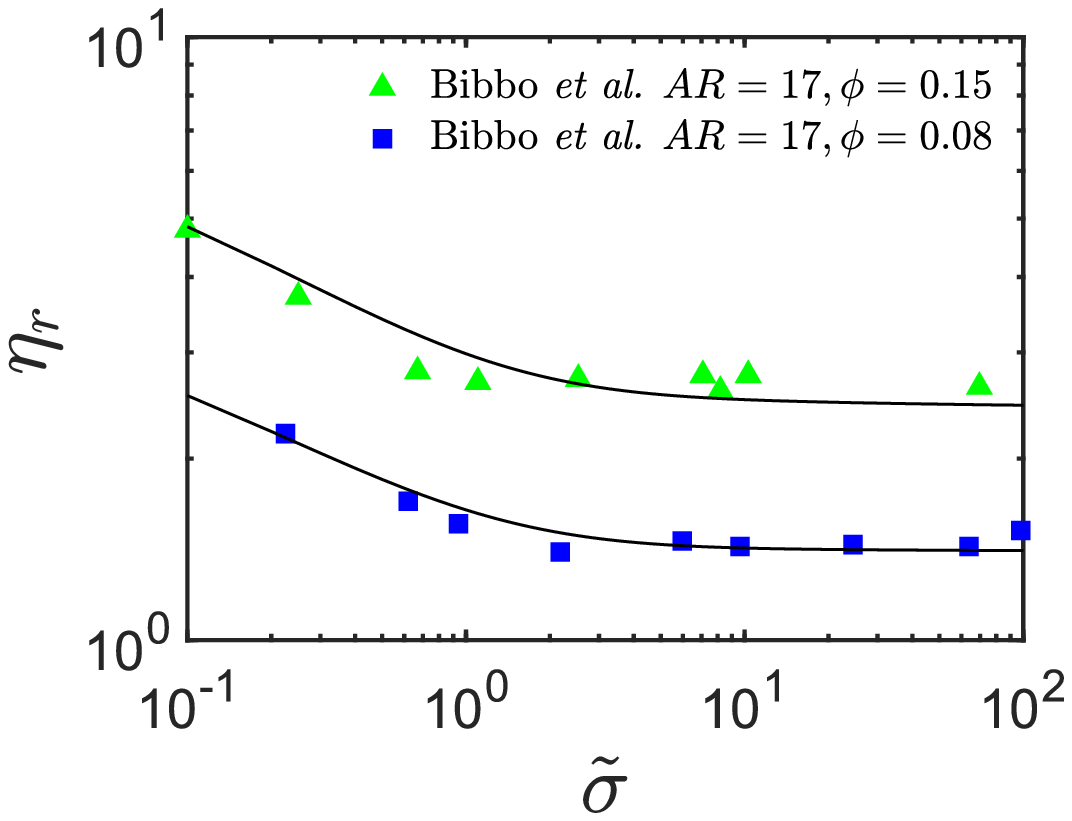}
  \caption{}
  \label{fig:stress_com}
\end{subfigure}%zz
\begin{subfigure}{.5\textwidth}
  \centering
  \includegraphics[width=1.0\linewidth]{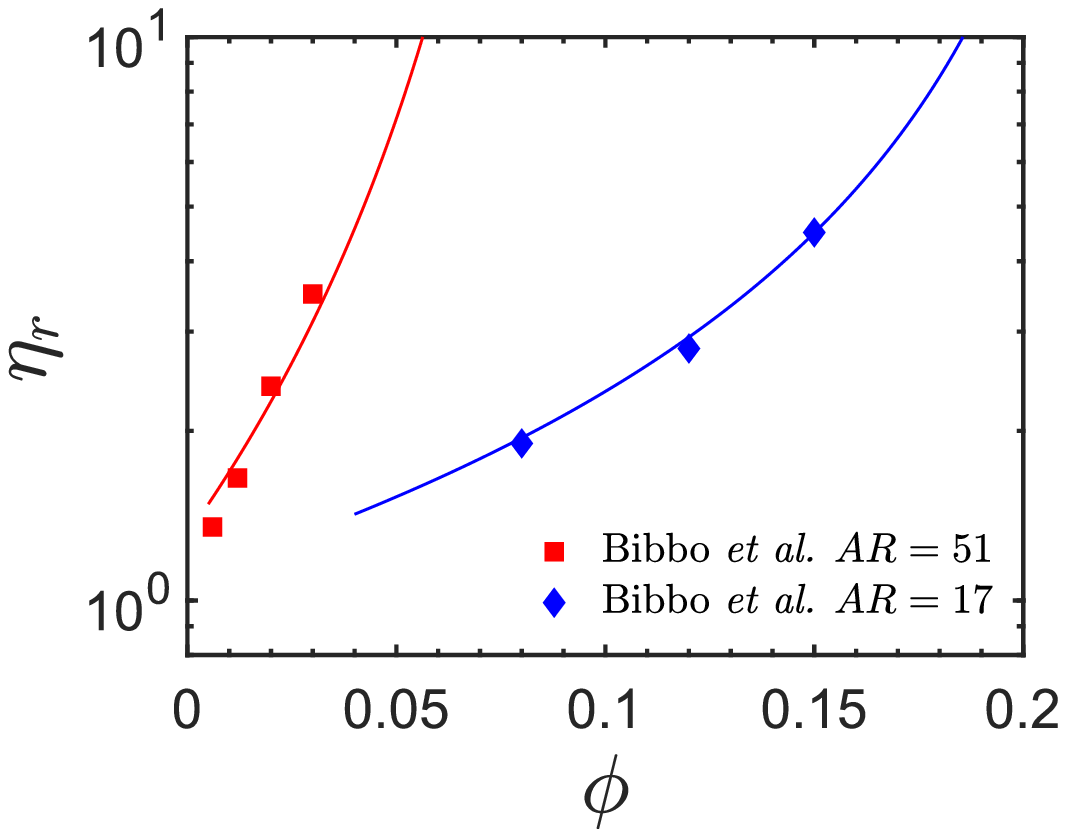}
  \caption{}
  \label{fig:jamming_com}
\end{subfigure}
\caption{Relative viscosity as a function of dimensionless shear stress, $\tilde{\sigma}$: (a) for fiber aspect ratio, $AR=10$ and (b) for fiber aspect ratio, $AR=36$. (c) \& (d) Comparison of the constitutive model with the experimental data from Bibbo \textit{et al.} \citep{bibbo1987rheology}. The shear stress was fixed to  $\tilde{\sigma} = 0.02$ when comparing the volume fraction-dependent relative viscosity with the experimental data.
The solid lines represent Eq.~\ref{eq:viscosity} with the values of the calibrated model parameters obtained using the simulation data.}.
\label{fig:viscosity_stress_AR}
\end{figure*}

Figure~\ref{fig:viscosity_stress_AR} shows the shear rate dependent relative viscosity for the short ($AR = 10$) and long ($AR=36$) fiber suspensions for the volume fractions investigated in the study. %As we present for $AR = 10$ and $AR = 36$ in the main text, we refer to the cases as "short" and "long" henceforth in the article.
The proposed model accurately predicts relative viscosity in the intermediate stress ($\tilde{\sigma}$) regime for both aspect ratios. 
It follows from this model that the reduced force that controls friction, and consequently the suspension relative viscosity, is proportional to the reduced shear stress: 
\begin{equation}
    \frac{F_n}{F_0}=\frac{\pi\eta\eta_rd^2\dot\gamma}{20F_0}=\frac{\eta_r(\dot\gamma/\dot\gamma_0)}{20}=\frac{\tilde\sigma}{20}
\end{equation}

We can use the values of calibrated parameters obtained from the simulation data to predict the stress-dependent relative viscosity of fiber suspensions with aspect ratio 17.  The prediction is compared to the experimental data from \cite{bibbo1987rheology}, as shown in figure~\ref{fig:stress_com}. The model does a satisfactory job of capturing the relative viscosity of the experimental system. For this comparison, we assume $\sigma_0 = 0.45$ to non-dimensionalize the experimental stress. Finally, we predict the volume fraction-dependent viscosity at a high shear rate for aspect ratios 17 and 51 and compare it with \cite{bibbo1987rheology}, as shown in figure~\ref{fig:jamming_com}. The data was only available in the semi-concentrated regime, and our simulation did a good job of capturing the experimental data. 

Finally, due to the agreement between the results from the load-dependent friction simulations and the viscosity from Eq.~\ref{eq:viscosity}, it is possible to estimate the effective coefficient of friction, $\mu_{eff}$, in the experiment without running new simulations. Once we know the jamming volume fraction from the experiments, $\mu_{eff}$ can be computed by reversing Eq.~\ref{eq:constant_fric_mu_alpha}. In this case, $\mu_{eff}$ is actually the microscopic friction coefficient for the applied load, $F_n=\pi*d^2\sigma/20$, where $\sigma$ is the experimental stress. For the reported experiment of Bibbo \textit{et al.} \citep{bibbo1987rheology}, we find the coefficient friction to be 1.12. We note that it is possible to deduce $\mu_{eff}$ from the viscosity measurements only if the values of the  experimental  jamming  fraction  belong  to  the variation range  of $\phi_m$, deduced  from the  numerical simulations  at  constant  friction  coefficient. It is important to remember that the friction coefficient values derived from the viscosity measurements are estimates.

\subsection{Flow state diagram}
\begin{figure}
\centering
\includegraphics[width=1.0\linewidth]{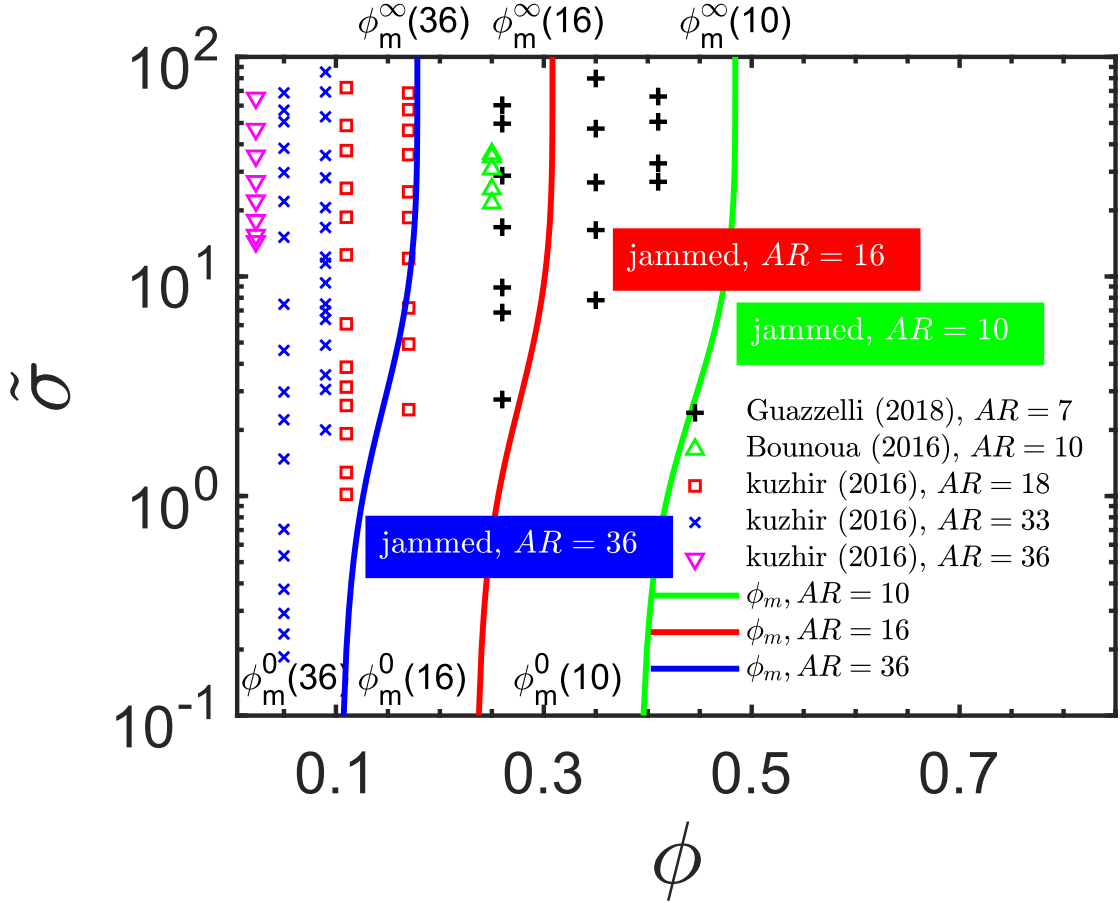}
\caption{$\tilde{\sigma}-\phi$ phase space diagram for fibers $AR = 10, 16$, and 36. Increasing aspect ratio decreases the jamming volume fraction.  The suspension is jammed for the region on the right of the solid curve.}. \label{fig:flow_state}
\end{figure}

The shear rheology described above is controlled by three dimensionless parameters, namely, dimensionless shear stress $\tilde{\sigma}$, volume faction $\phi$, and aspect ratio $AR$. The results discussed in the paper are presented in a flow state diagram in figure~\ref{fig:flow_state}. Here in the $\tilde{\sigma}-\phi$ phase space, we identify  $\phi_m^\infty$, $\phi_m^0$, and $\phi_m (\tilde{\sigma})$. In the lower part of the diagram, when the stress is too low, and in the upper part of the diagram, when the stress is large, rheology diverges at $\phi_m^0$, and $\phi_m^\infty$, respectively. So in the two extremes, the relative viscosity is rate-independent. In between, we observe rate-dependent viscosity.

Moreover, the volume fraction at which the suspension jams increases with increasing stress. Previous studies on the suspension of fibers have reported that at low stress values, the suspension does not flow but can flow at higher stress values \citep{bibbo1987rheology, chaouche2001rheology, switzer2003rheology, keshtkar2009rheological,bounoua2016apparent,bounoua2016normal} meaning that the jamming volume fraction, $\phi_m$ depends on the shear stress. However, the exact dependence of the jamming fraction on the stress was largely unexplored. Our numerical simulation is superior in the sense that it quantifies the exact dependence of the jamming volume fraction on the applied stress denoted by the solid line in figure~\ref{fig:flow_state}. In addition, the applicability of the model is further strengthened  in capturing the dependence of jamming volume fraction on the stress for different aspect ratios. 

%$AR = 10, 16$, and 36.

\section{Conclusion}

This paper presents a constitutive model for frictional fiber suspensions in a steady shear flow. The proposed model quantifies the effects of three independent parameters - fiber volume fraction $\phi$, dimensionless stress $\tilde{\sigma}$, and fiber aspect ratio $AR$. The model quantitatively predicts the viscosity of a  range of aspect ratios and volume fractions once it is calibrated using the zero and high shear viscosities of just a few aspect ratio cases.

%The model requires knowledge of rheology at zero and high shear rate limits for a few aspect ratios and the rheology in between can then be interpolated using the equations presented here. 

%The viscosity model proposed here shows that by considering load dependent variable friction, it is possible to capture the shear thinning rheology observed in the most concentrated fiber suspension. In this model, the effective friction coefficient is described by the effective normal contact force, which is directly proportional to the shear stress. As the shear stress increases, the  effective friction coefficient reduces, and the jamming volume fraction rises, resulting in a decrease in relative viscosity. The model and the simulation results have a very good agreement. 

We start our analysis by deriving relations describing the dependence of jamming volume fraction on the friction coefficient to reflect the impact of this microscopic parameter on the suspension rheology. However, constant friction cannot explain the rate-dependent rheology in suspensions. Hence, we use the relations derived from the constant coefficient of friction for quantitatively comprehending the case with load-dependent friction coefficient.  This approach assists us in defining the stress-dependent jamming fraction $\phi_m(\tilde{\sigma})$ by interpolating between two jamming fractions at the extreme limit of stresses, in the manner proposed by Wyart and Cates \citep{wyart2014discontinuous} for shear thickening suspensions. The divergence of viscosity approaching  the interpolated jamming volume fraction follows the form of the two limits, with the viscosity growing as $\phi_m^{-0.90}(\phi_m-\phi)^{-0.90}$. Furthermore, we extend the constitutive model to capture the aspect ratio-dependent rheological behavior of the suspension. Once the aspect ratio-dependent rheological properties in the low and high shear limits are known, we can use the constitutive model to quantify the aspect ratio's effect on the rheology between the two stress limits. The model predictions for the relative viscosity, $\eta_r(\phi,\tilde{\sigma},AR)$ agree well not only over the full range of parameters it is calibrated on but beyond this range as well as evident from a good agreement between experimental measurements and model predictions. In the end, we display the dependence of the jamming volume on the applied stress and aspect ratio of the fibers through a flow state diagram in the $\phi - \tilde{\sigma}$ plane.

Finally, these results can be used to quantitatively predict the shear-thinning suspension behavior at different fiber aspect ratios and volume fractions. This model can be used to estimate the effective coefficient of friction in the experiment from the jamming volume fraction. The idea thus gained can be utilized in manipulating suspension behavior by changing the fiber size along with mechanisms based on hydrodynamic
interactions, fiber surface roughness \citep{khan2021rheology}, and friction \citep{salahuddin2013study, switzer2003rheology}. We found that to have higher solid concentrations desired in industrial applications, we should break down the fibers into smaller sizes or reduce the coefficient of friction through surface treatment.

\begin{acknowledgments}
AMA would like to acknowledge financial support from the Department of Energy via grant EE0008910. 
\end{acknowledgments}

%\nocite{*}
\bibliography{paper}% Produces the bibliography via BibTeX.

\end{document}